# Extreme UV generation from molecules in intense Ti:Sapphire light

**J F McCann, L-Y Peng, I D Williams, K T Taylor, D Dundas**

*School of Mathematics and Physics, Queen's University Belfast, Belfast BT7 1NN, Northern Ireland, UK*

**Main contact email address:** *j.f.mccann@qub.ac.uk*

## Introduction

The use of ultrashort intense infrared light to study atomic and molecular dynamics on femtosecond timescales has been extremely successful. This technique, particularly in the pump-probe mode, has enabled an analysis in the time domain of molecular conformation changes and the study of excited-state vibrational dynamics. Very recently the prospect of resolving and manipulating electronic dynamics by creating high-frequency optical pulses has been advanced. To realise spatial and temporal resolution of electronic transitions within a molecule requires intense coherent light that switches on attosecond ($10^{-18}$s) timescales with wavelengths in the extreme ultraviolet or soft x-ray spectrum, $\lambda=0.1-10$ nm.

One of the most promising schemes for producing such light is through high-harmonic generation from intense Ti:Sapphire lasers interacting with gas-phase atoms and molecules[1]. The CCLRC-Central Laser Facility, Imperial College and Oxford are leading participants in a large UK consortium developing sub-femtosecond and attosecond pulses using the controlled generation of high-frequency light in atomic and molecular systems. The process of high-harmonic generation from IR light can be considered to have three distinct steps. Firstly, the initial molecular orbital is polarized and pulled away from the parent core by the external field to form a coherent continuum state, which then evolves within the laser field and finally, a few femtoseconds later driven by the field reversal, it returns to interact with the molecular core. In the ensuing collision the electron releases energy by bremsstrahlung during its passage across the molecule. Experiments have shown that the nonlinear medium, the molecule, has a critical role in forming, evolving and scattering the electronic wave packet, and hence in the generation of extreme UV light[2].

A feature of molecules that are hydrogenated or deuterated is that, for IR stimulation, the hydrogen vibrational relaxation time is comparable with the cycle time of the field, and hence the excursion time of the continuum wave packet. In previous work, we have shown that this vibration is extremely important for the process of ionization in light molecules[3] and in this work we show that it carries over to the process of harmonic generation. We study the quantal vibration effect in both long and short pulses and for both isotopes of hydrogen. In order to simplify the physics and isolate the important mechanisms, we consider the simple one-electron molecular ion and the single-molecule response for the spectral density.

## Quantum vibration effects in 25 fs pulses

We simulate the dynamics of $H_2^+$ and $HD^+$ by direct solution of the time-dependent Schrödinger equation for the electronic and nuclear motion for the interaction of intense femtosecond pulses. On these timescales the rotational motion, even for such light molecules, is frozen. Therefore it is a reasonable assumption that the nuclear alignment is fixed during the pulse interaction and that rotation can be neglected. In terms of vibrational relaxation, and since the nuclei are light, vibration will be important over femtosecond timescales. Although homonuclear diatomics are IR-inactive, in an intense field one can create vibrational excitation through continuum coupling. To show the effect of vibration, consider a first approximation in which the nuclei are infinitely massive so they maintain their positions at a fixed bond length of R=2 a.u., throughout the process. A typical result for the single-molecule response for a 10 cycle, $\lambda=750$nm pulse is shown in Figure 1. The spectrum has a characteristic plateau that is in rough agreement with the classical cut-off. In the classical model the maximum photon energy depends on the ionization potential of the molecule ($I_0=31.7$ eV) and ponderomotive energy corresponding to an intensity of $I=4.5 \times 10^{14}$ W cm$^{-2}$ ($U_P=23.6$ eV). In this case, and under the assumption of fixed nuclei, only 1% of the molecules are ionized. A more realistic simulation, including quantal vibration, shows that 38% of the $H_2^+$ molecules are ionized. For the heavier isotope, $HD^+$ the yield is lower at 26% and closer to the infinite mass approximation. Our simulation shows that quantum vibration gives dramatically different results for ionization yields.

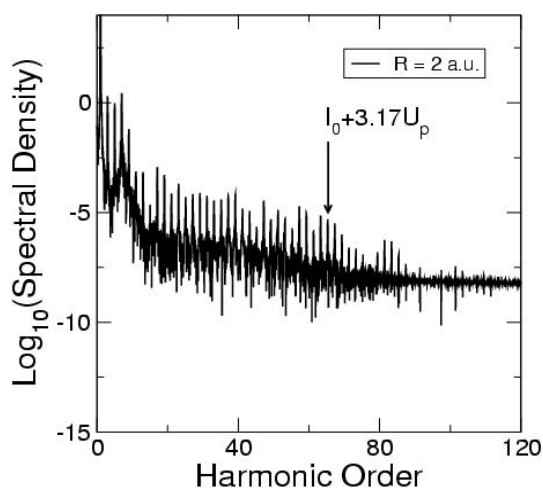

**Figure 1**. Simulation for frozen vibration showing the harmonic generation spectrum from the hydrogen molecular ion with fixed molecular bond length (R=2 a.u.). Indicated on the Figure is the classical cut-off frequency, which depends on the ionization potential ($I_0$) and ponderomotive energy ($U_P$). The incident pulse is 10 cycles, $\lambda=750$nm, with intensity, $I=4.5 \times 10^{14}$ W cm$^{-2}$.

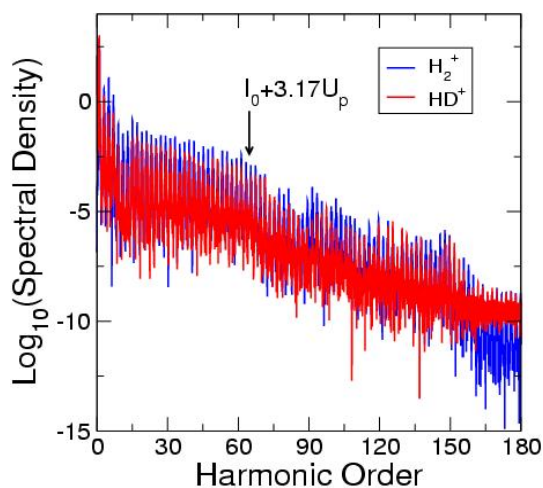

**Figure 2.** Simulation with quantal vibration. The harmonic generation spectrum from the hydrogen molecular ion including quantal vibration. The incident pulse is the same as that used for the results in Figure 1.

The corresponding harmonic generation spectrum including quantum vibration for the same 10 cycle, λ=750nm pulse with intensity, I=4.5x10$^{14}$ W cm$^{-2}$, is shown in Figure 2. The spectrum of Figure 1 shows very significant inadequacies in the infinite mass approximation in qualitative and quantitative terms when compared with the quantum simulation (Figure 2). The molecule is much more efficient at high harmonic generation than one would conclude based on the infinite mass approximation. For example, around the 40$^{th}$ harmonic the quantum vibration amplifies the intensity by a factor of 100 in comparison with Figure 1. Furthermore, the classical cut-off law is no longer a reliable guide to the range of the plateau in Figure 2. Indeed the spectral density shows strong yields around 5nm, corresponding to the 150$^{th}$ harmonic, not replicated by Figure 1.

The primary reason for these quantal features is that molecular vibration assists the ionization process occurring via dynamic tunnelling ionization[3]. One could cite this as an extreme example of phonon-assisted tunneling. Another factor is the quantum enhancement of bremsstrahlung due to nuclear wave packet expansion and dispersion increasing the cross section for electron-ion scattering.

Additional quantal effects are visible in the structure of the spectral lines. A closer analysis of the spectrum, see Figure 3 upper diagram, around the 45$^{th}$ harmonic shows that the sidebands of the triplet are prominent and well resolved. This is further evidence of strong continuum coupling. Conversely the fixed nuclei results show sharp central lines with very weak sidebands. We note that the small permanent dipole moment of HD$^+$ gives even harmonics for all harmonics. Around the 85$^{th}$ harmonic (Figure 3 lower) the even harmonics are present for both species indicating polarization during the pulse rise – a feature of high intensity and short pulse rise time.

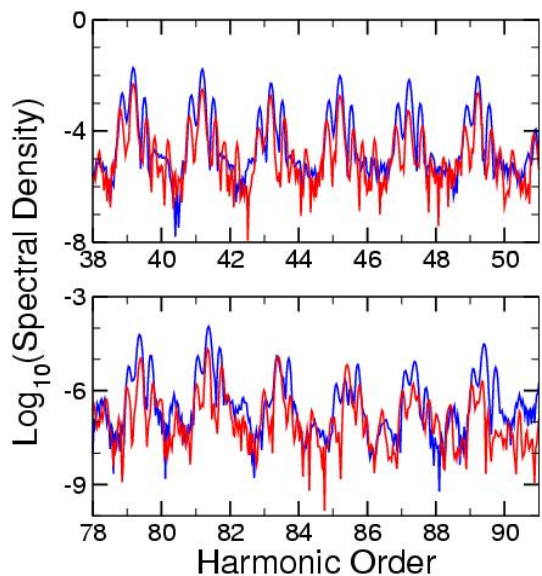

**Figure 3.** Simulation with quantal vibration. Expanded from Figure 2. The incident pulse is the same as that used for the results in Figure 1.

**Quantum spectra for 5 fs pulses**

Recent advances have led to the production of IR pulses that contain only a few optical cycles[1]. In Figure 4 we show full quantal simulations of few cycle intense IR pulses with molecular systems for λ=750nm, and I=1x10$^{15}$ W cm$^{-2}$.

Although the intensity is slightly higher than before (Figure 2), we do not saturate the process. For this case we find that 15% of the H$_2^+$ molecules are ionized and 12% of the HD$^+$, showing the less important role of inertia in the ionization process. As for harmonic generation, in Figure 4 we note that the classical cut-off reappears – again a feature of the weaker effect of vibration.

The low-order spectral lines are broader than those of Figure 2 due to the bandwidth of the incident pulse and there is some significant blue shifting. In terms of efficiency, the yield around the 100$^{th}$ harmonic is comparable with the longer, lower-intensity pulse (Figure 2). However extreme UV generation is strongly attenuated and limited by I$_0$. Another feature of this short pulse is unusual resonance lines at the lower harmonics.

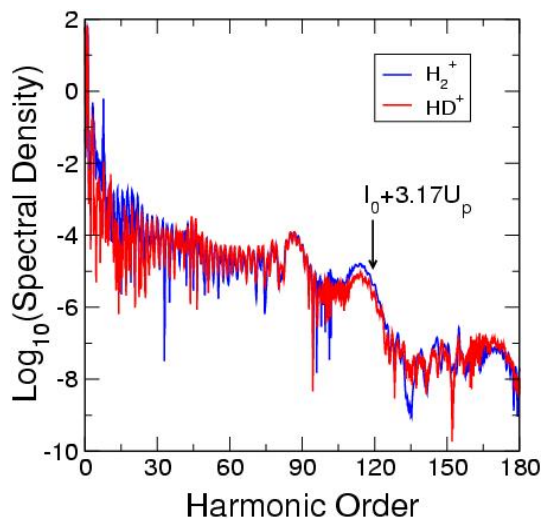

**Figure 4.** Simulation with quantal vibration as in Figure 2. The incident pulse has a Gaussian envelope FWHM 5fs, λ=750nm, with intensity, I=1x10$^{15}$ W cm$^{-2}$.

**Conclusions**

The simulations show that quantal vibration enhances the high-harmonic yield from hydrogen and deuterium molecules for intense femtosecond IR pulses. The spectral line shapes, sidebands and intensities of the harmonics show quantum features and can be calculated to high precision for arbitrary IR pulses. This will provide a useful tool for attosecond pulse design. For few-cycle IR pulses the molecular expansion is suppressed and the classical cut-off formula gives a reasonable limit for the harmonic plateau. The yields (intensities) of extreme UV light from these light dimers would be rather low in comparison with heavier multi-electron systems, although the a smaller ionization potential will limit the extent of the plateau. Diatomic molecules such as O$_2$ and N$_2$ relax slowly and thus quantal vibration will have a minor role for high harmonic generation using fs pulses. However IR-active molecules such as CO$_2$ will undoubtedly show quantal rovibration features of the initial state that will be an important factor in the spectrum and hence of interest to future developments in this field.

**Acknowledgements**

Thanks are due to EPSRC for provision of computer resources at CSAR that made these calculations possible. We are also very grateful to Prof J Marangos for useful discussions on the applications of this work. Mr L-Y Peng is grateful to the IRCEP at Queen's University for financial support.